\newcommand \be{\begin{equation}}
\newcommand \ba{\begin{eqnarray}}
\newcommand \ea{\end{eqnarray}}
\newcommand \ee{\end{equation}}

\documentclass[twocolumn,preprintnumbers,amsmath,amssymb]{revtex4}
\usepackage{color}
\usepackage{graphicx}
\usepackage{graphics}

\begin{document}

\tolerance=10000
\title{Transmission properties of optical adhesives and bonding layers }

\author{Arsen Subashiev$^*$ and Serge Luryi}
\affiliation{Department of Electrical and Computer Engineering,
State University of New York at Stony Brook, Stony Brook, NY,
11794-2350}
\email{subashiev@ece.sunysb.edu}

\begin{abstract}
We analyze the transparency of a thin film of low refractive index
(an optical glue or a bonding layer) placed between higher-index
media and forming an opto-pair.  Examples include a semiconductor
light-emitting diode with attached lens or a semiconductor
scintillator bonded to a photodiode. The transparency of an
opto-pair is highly sensitive to the film thickness due to the
so-called frustrated total internal reflection. We show that high
transparency in a wide range of the incidence angle can be achieved
only with very thin layers, more than an order of magnitude thinner
than the wavelength. The angular dependence of the transmission
coefficient is shown to satisfy a simple and universal sum rule.
Special attention is paid to the angular average of the optical
power transmission, which can be cast in a universal form for two
practically relevant classes of source layers.
\end{abstract}
%\ocis{230.0250, 310.6870}
\maketitle

\section*{Introduction}
A number of semiconductor optoelectronic devices require optical
matching of their component layers.  Semiconductor light emitting
diodes (LED) have low external emission efficiency limited to 2-4 \%
due to the high refractive index of the source crystal, resulting in
a narrow escape angle of total internal reflection (TIR)
\cite{Moss}. To avoid the TIR, various techniques have been used
including random surface texturing, pyramidal-shaped structures, as
well as  devices exploiting ``wave optics" effects and optimizing
interference  in the resonant cavity (see \cite{Delbeke} for the
review). The most effective remedies for the TIR effects in LEDs are
optically tight lenses, resulting in external emission efficiencies
approaching unity \cite{{LED1}, {LED2}}.  Another example when
optical matching matters is  a semiconductor scintillator used for
the detection of high-energy particles with the scintillating
radiation registered by a photodiode \cite{Serge}. In both cases the
optical components can be attached by using optical adhesives or
optical glues in the form of a thin interlayer. The refractive index
of the interlayer is considerably lower than that of both
semiconductors. It is usually expected that films thinner than
wavelength should not disturb the radiation transmission or
waveguiding properties \cite{Bowers}. This is indeed true for normal
incidence. Moreover, for a fixed angle of incidence the transparency
can be generally enhanced by choosing the thickness of the low-index
layer so as to provide constructive interference for the transmitted
wave. However, for an isotropic source of radiation (as in a LED or
a scintillator) the angles of incidence are widely spread and a
considerable part of incident radiation is in the TIR region for the
interlayer interface. For this part of radiation, the transmission
is provided by evanescent optical waves and is exponentially small.
The residual transparency is known as the {\it frustrated} total
internal reflection (FTIR) \cite{Zhu}. In the FTIR region the
transparency remains dependent on the layer index matching, but the
requirements for the intermediate layers become much more
restrictive.

In this paper, we first analyze the transparency coefficient of a
thin low-index film for arbitrary incident angles and polarization.
We then calculate the average optical power transmission of
isotropic radiation and discuss the stringent requirements on the
interlayer thickness.

\section*{Transparency of a thin film with frustrated total internal reflection}
\subsection*{General analysis}
\begin{figure}[b]
\centerline{\includegraphics[width=0.45\textwidth,clip]{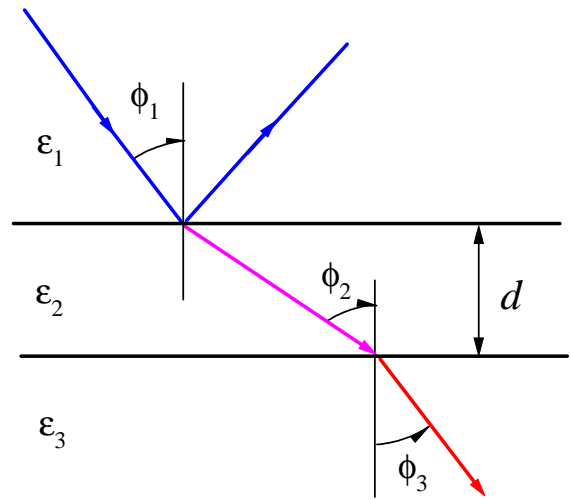}}
\caption[] {(Color online) Reflection and transmission of a wave
incident on a homogeneous layer.} \label{Fig0}
\end{figure}

First, we consider the reflection of light with wavelength $\lambda$
incident on a layer of thickness $d$ from a material of dielectric
constant  $\epsilon_1=n_1^2$ for a fixed angle of incidence
$\phi_1$, as shown in Fig. \ref{Fig0}.
%%%%%%%%-------------------------------
We assume that the permittivity of the intermediate layer
$\epsilon_2=n_2^2$ is real (negligible absorption) and smaller than
that of the surrounding media on both sides. The 3rd layer
permittivity is $\epsilon_3=n_3^2$. While we take $\epsilon_2 <
\epsilon_1$ and $\epsilon_2 < \epsilon_3$, no special relation
between $\epsilon_1$ and $\epsilon_3$ is assumed. An exemplary
structure we consider is a pair of semiconductor plates, optically
coupled using an optical glue \cite{LED1} or a bonding oxide
\cite{Bowers}.

The solution for the amplitude reflectance, first obtained in
\cite{Foester} (see also e.g. \cite{BornWol}, or \cite{L&L}), is of
the form \be r= \frac{r_{12} + r_{23} \exp
(2i\delta)]}{1+r_{12}r_{23} \exp (2i\delta)}~. \label{ref0} \ee Here
$r_{12}$ and $r_{23}$ are the Fresnel reflection coefficients at the
interfaces 1-2 and 2-3, respectively, and $\delta$ is the phase
shift of the wave inside the film, \be \delta=\frac{2\pi d
}{\lambda}\sqrt{\epsilon_2-\epsilon_1\sin^2 \phi_1}~. \label{delta}
\ee For a large $d$ one can define the total reflection angle
$\phi_t$ by $n_1\sin \phi_t  = n_2$, which corresponds to the TIR
for incident angles $\phi_1\ge \phi_t$. When the phase shift is
small, $\delta \ll 1$, the film is called {\it optically thin}; we
note that close to $\phi_t$ the film is always optically thin.

The amplitude  reflectance depends (through the reflection
coefficients $r_{12}$ and $r_{23}$) on the polarization of light,
which will be indicated below (when necessary) by an additional
subscript. We shall use subscript $s$ for $s$-polarized waves
(electric field perpendicular to the plane of incidence) and $p$ for
$p$-polarization (electric field in the plane of incidence).

The Fresnel reflection coefficients for the two interfaces are given
by 
\ba  r_{ij,s}= \frac{n_i\cos \phi_i -n_j\cos \phi_j }{n_i\cos
\phi_i +n_j\cos \phi_j }~,\nonumber  \\ r_{ij,p}= \frac{n_j\cos
\phi_i -n_i\cos \phi_j }{n_j\cos \phi_i +n_i\cos \phi_j}~.
\label{ref12}
\ea
where $i$ and $j$ are interface indices:
$\{ij=12\}$ for the 1-2 interface  and $\{ij=23\}$ for the 2-3
interface, respectively. Note that $r_{ij}=-r_{ji}$  \cite{ref}.

Using Eqs. (\ref{ref0}), (\ref{delta}), and (\ref{ref12}), we
calculate the reflection coefficient $R=|r|^2$. The transmission
coefficient for non-absorbing layer is then readily obtained as
$T=1-R$.

For small angles of incidence,  $\phi_1 \le \phi_t$, the phase shift
$\delta$ and the interface reflection coefficients for both
polarizations are real numbers (moreover, $r_{32}>0$), so that $
R_{ij}=r_{ij}^2$ and we have \be R= \frac{(r_{12}-r_{32})^2 +
4r_{12} r_{32}\sin^2 \delta }{(1-r_{12} r_{32})^2+4r_{12}
r_{32}\sin^2 \delta }~. \label{rfin} \ee Equation (\ref{rfin}) can
be simplified using Fresnel's equations and Snell's refraction law,
$n_1 \sin\phi_1=n_2\sin \phi_2=n_3\sin\phi_3$. After some algebra,
the transmission coefficient can be brought into the form \be T=
\frac{T_{13}}{1+aT_{13}\sin^2\delta}~, \label{Tfin} \ee where
$T_{13}$ is the transmission coefficient of the interface 1-3. For a
non-zero angle of incidence  both $T_{13}$ and the coefficient $a$
are different for the two polarizations.

For $s$-polarization we have: \be
a_s=\frac{(n_{12}^2-1)(n_{32}^2-1)}{4n_{12}^2 \cos \phi_1
(1-n_{12}^2\sin^2 \phi_1 )\sqrt{n_{31}^2-\sin^2 \phi_1 }}~,
\label{as}\ee and \be T_{13,s}= \frac{4\cos \phi_1
\sqrt{n_{31}^2-\sin^2 \phi_1 }}{(\sqrt{n_{31}^2-\sin^2 \phi_1 }+\cos
\phi_1 )^2}~. \label{bs}\ee Here and below we use the notation
$n_{ij}=n_i/n_j$.

For $p$-polarization we find: \be
a_p=\frac{a_s}{n_{31}^2}[(n_{12}^2+1) \sin^2 \phi_1
-1][(n_{32}^2+1)\sin^2 \phi_1 -n_{31}^2]~, \label{ap}\ee and \be
T_{13,p}= \frac{4n_{31}^2\cos \phi_1 \sqrt{n_{31}^2-\sin^2
\phi_1}}{(\sqrt{n_{31}^2-\sin^2 \phi_1}+n_{31}^2\cos \phi_1 )^2}~.
\label{bp}\ee

Using some caution, one can apply Eq. (\ref{Tfin}) in the full range
of variation of the angle of incidence, including the FTIR region
($\phi_1 > \phi_t$). In the FTIR region, the phase gain $\delta$
becomes imaginary, $\delta=i\delta'$, where \be \delta'= \frac{2\pi
d n_2}{\lambda}\sqrt{n_{12}\sin^2 \phi_1 -1}~, \label{deltaprim} \ee
while the reflection amplitudes become unimodular,
$r_{12}=\exp(i\delta_{12})$. The reflection phase $\delta_{12}$ and
other relevant phases can be readily written down using Fresnel's
equations. In the range of FTIR, the transmission coefficient is
then given by \cite{CourtWil} \be T= \frac{T_{13}}{1-aT_{13}\sinh^2
\delta'}~, \label{Tfinf} \ee where both $T_{13}$ and $a$ are given
by the same Eqs. (\ref{as}-\ref{bp}) but $a_s$ and $a_p$ become
negative.

Note that the nature of reflection in the FTIR region,
$\phi_1>\phi_t$, is quite different, since the reflection
coefficient from a single surface should be equal to unity. The
complete single-surface reflection is frustrated by the interference
with light reflected by the second surface, even though the field
between the two surfaces has a totally evanescent character. For
$\phi_1=\phi_t$ the optical field deeply penetrates into the
interlayer. If $n_{31}< 1$, the transmission coefficient $T$
vanishes (and so does $T_{13}$) for  $ \sin\phi_1=n_{31}$, due to
the TIR from layer 3.

%%%%%%%%%%%%%%-------------------------
\begin{figure*}[t]
\centerline{\includegraphics[width=0.86\textwidth,clip]{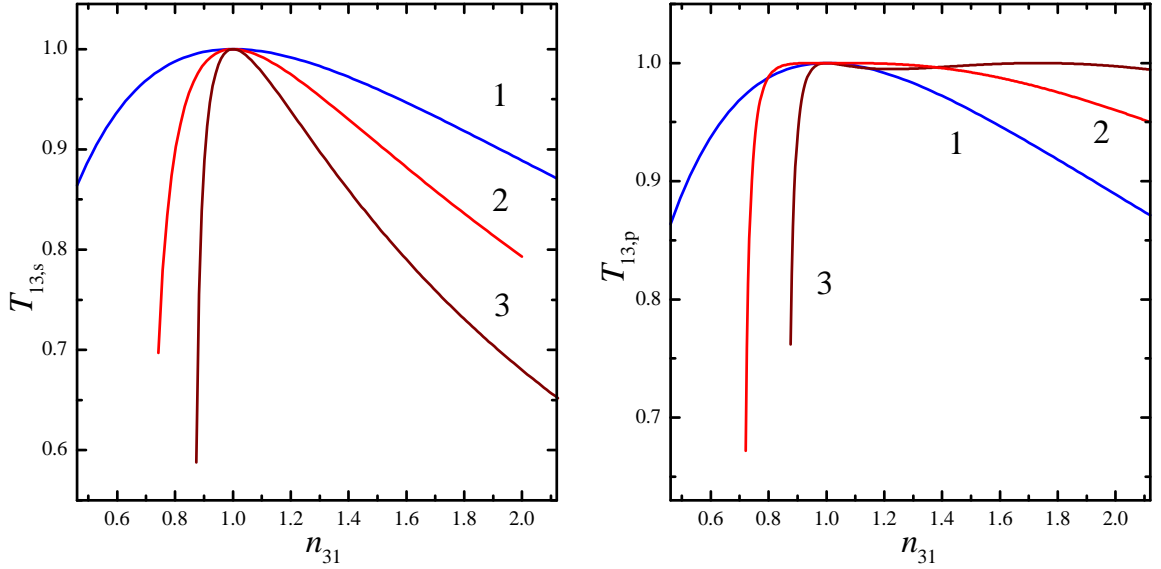}}  \caption[]
{(Color online) Interface transmission coefficients $T_{13,s}$ (left
panel) and $T_{13,p}$ (right panel) as functions of $n_{31}=n_3/n_1$
for several values of the incidence angle and two polarizations of
the incident wave. (1): $\cos \phi_1=$0; (2): $\cos \phi_1=$0.5; and
(3): $\cos \phi_1=$0.7.} \label{Fig1b}
\end{figure*}
%%%%%%%%%%%%%%%-------------------------

According to  Eqs. (\ref{Tfin}) and (\ref{Tfinf}), for a
sufficiently thin film ($\delta \ll 1$ and $\delta' \ll 1$) the
transmission coefficients approach $T_{13,s}$, $T_{13,p}$. These
values in turn depend on $n_{31}$ and $\phi_1$. The dependencies are
presented in Fig. \ref{Fig1b}. For all incident angles the total
transparency, $T_{13,s}=T_{13,p}=1$, is reached only when
$n_{31}=1$. Away from the exact index matching the decrease of
$T_{13,s}$ and $T_{13,p}$ is seen to be much steeper on the side
$n_{31}\le 1$. Since the exact matching of indices of the layers 1
and 3 is rarely possible, the structures with $n_{31} \ge 1$ are
preferable.

For small incidence angles, $\phi_1<\phi_t$, one has a constructive
interference between the waves in the middle layer, which results in
the maximum of the transmission coefficient for both polarizations
at $\delta=\pi m$, where $m$ is an integer, $m=1,2 ...$ (for normal
incidence this corresponds to $d= m \lambda/2n_2$). Note that this
anti-reflection effect arises for half-wave plates
--- rather than quarter-wave plates, which would be the case for an intermediate
layer in structures with  $n_1<n_2<n_3$ or  $n_1>n_2>n_3$. The
almost total transparency of the film at $m=0,1,2 ...$ is due to the
phase shift in the reflection from the two film surfaces, which
results in $r_{12} \approx -r_{23}$. The transmission coefficients
in these interference maxima are given by $T_{13,s}$ and $T_{13,p}$,
Eqs. (\ref{bs}) and (\ref{bp}), respectively.

For normal incidence ($\phi_1=0$) one has
$T_{13,s}=T_{13,p}=T_{13}$, where 
\be
T_{13}=\frac{4n_{31}}{(1+n_{31})^2}~, \hspace{1cm}
R_{13}=\frac{(1-n_{31})^2}{(1+n_{31})^2}~. \label{rmin} \ee For most
semiconductor pairs, the ratio $n_{31}$ is close to unity. For
typical pairs with indices in the range of 3.5 to 4, the reflection
losses at normal incidence are less than 0.5 \%.

For an optically thin film, $\delta_0 \ll1$, one has \be
T|_{\phi_1=0}=T_{13}-4T_{13}^2
\frac{\sqrt{R_{12}R_{23}}}{T_{12}T_{23}}~(2\pi n_2 \tilde d)^2 ~,
\label{rnormth} \ee where $\tilde d=d/\lambda$ and the reflection
and transmission coefficients for the 1-2 and 2-3 interfaces are
defined similar to $R_{13}$, $T_{13}$ in Eq. (\ref{rmin}). We see
that the deviation $T-T_{13} \propto  \tilde d^2$.
%%%%%%%%%%%%%%-------------------------
\begin{figure*}[t]
\centerline{\includegraphics[width=0.86\textwidth,clip]{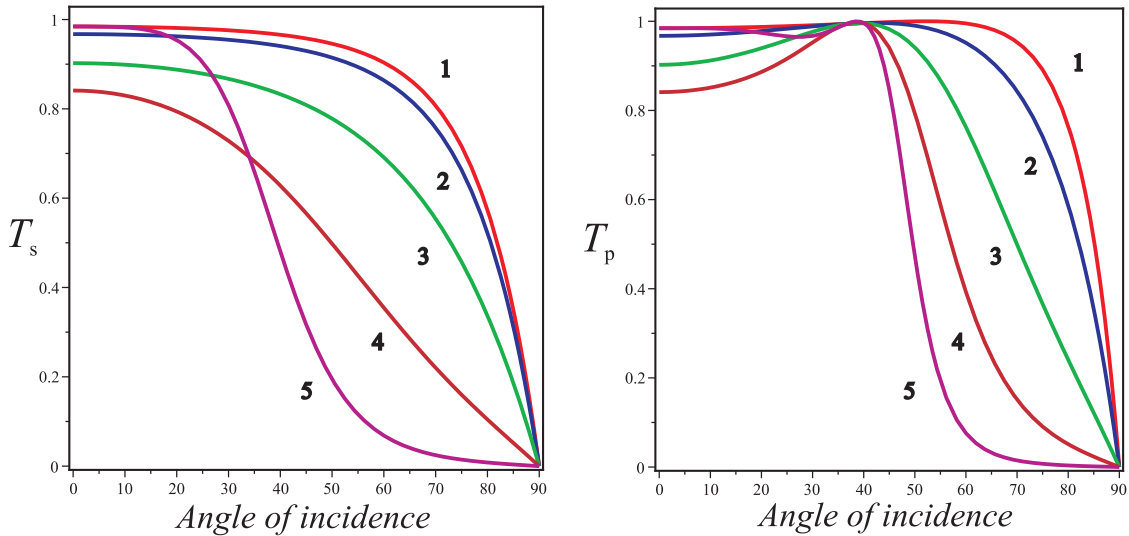}}
\caption[] {(Color online) Angular dependence of the transmission
coefficients $T_s$ and $T_p$ for a thin layer of index $n_2=2.6$;
the incident radiation from layer 1 with $n_1=3.5$ is transmitted
into layer 3 with $n_3=4.5$. Different curves correspond to
different film thicknesses described by a dimensionless parameter $
\tilde d=d /\lambda$, viz. (1): $\tilde d=$ 0; (2): $\tilde d=$
0.02; 3: $\tilde d=$ 0.05; 4: $\tilde d=$ 0.2, and 5: $\tilde
d=1/(2n_2)$. The dependence on the angle of incidence is displayed
in the range $0 < \phi_1 <\pi /2$ for two polarizations of incident
light: $s$-polarization (left panel) and $p$-polarization (right
panel). The TIR angle $\phi_t=$ 48$^\circ$, Brewster's angles are
$\phi_B=$ 36.6$^\circ$ for the 1-2 interface and $\phi_B'=$
40$^\circ$ for 2-3 surface.} \label{FigMLED}
\end{figure*}
%%%%%%%%%%%%%%%-------------------------
For $\phi_1 \rightarrow \phi_t$, the reflection coefficients go to
unity, $R_{12,s}=R_{12,p}=1$. This, however, does not imply $T=0$,
since one has $\delta \rightarrow 0$ at the same time. Therefore,
one needs a more refined consideration of the reflection in the
vicinity of $\phi_1=\phi_t$. It is convenient to give expressions
for the reciprocal transparencies in terms of the layer
permittivities, viz. 
\ba  \frac{1}{T_s}|_{\phi_1=\phi_t}=
\frac{(\sqrt{\epsilon_1-\epsilon_2}+\sqrt{\epsilon_3-\epsilon_2})^2}
{4\sqrt{\epsilon_1-\epsilon_2}\sqrt{\epsilon_3-\epsilon_2}}\nonumber \\ +\pi^2
\tilde d^2 \sqrt{\epsilon_1-\epsilon_2}\sqrt{\epsilon_3-\epsilon_2}
~, \label{Tlay-s} \ea
 and 
 \ba \frac{1}{T_p}|_{\phi_1=\phi_t}=
\frac{(\epsilon_3\sqrt{\epsilon_1-\epsilon_2} +\epsilon_1\sqrt{\epsilon_3-\epsilon_2})^2}
{4\epsilon_1\epsilon_3\sqrt{\epsilon_1-\epsilon_2}\sqrt{\epsilon_3-\epsilon_2}}\nonumber \\ +\pi^2
\tilde d^2 \sqrt{\epsilon_1-\epsilon_2}\sqrt{\epsilon_3-\epsilon_2}
~. \label{Tlay-p} \ea 
For matching  layers, $\epsilon_3=\epsilon_1$,
and $s$-polarized waves, we have 
\ba R_s |_{\phi_1=\phi_t}= \frac{
\pi^2 \tilde d^2 (\epsilon_1-\epsilon_2)}{1+\pi^2 \tilde d^2
(\epsilon_1-\epsilon_2)}, \nonumber \\  T_s|_{\phi_1=\phi_t}= \frac{
1}{1+\pi^2 \tilde d^2 (\epsilon_1-\epsilon_2)}~. \label{rbotot} \ea
Similarly, for matching layers and waves polarized in the plane of
incidence 
\ba R_p|_{\phi_1=\phi_t}= \frac{ \pi^2 \tilde d^2
(\epsilon_1-\epsilon_2)(\epsilon_2/\epsilon_1)^2}{1+\pi^2 \tilde d^2
(\epsilon_1-\epsilon_2)(\epsilon_2/\epsilon_1)^2} , \nonumber \\
T_p|_{\phi_1=\phi_t}= \frac{ 1}{1+\pi^2 \tilde d^2
(\epsilon_1-\epsilon_2)(\epsilon_2/\epsilon_1)^2}~. \label{rpartot}
\ea 
It follows from Eqs. (\ref{rnormth}), (\ref{rbotot}) and
(\ref{rpartot}) that (i) for perpendicular ($s$) polarization, the
reflection increases steadily as a function of the incidence angle,
but it remains small up to the TIR angle; and (ii) the reflection
for $s$ polarization is stronger than that for the parallel ($p$)
polarization. The reflection of $p$-polarized light is suppressed
due to Brewster's effect: for incidence $\phi_1$ at the Brewster
angle, $\phi_B=\arctan(n_2/n_1)$, one has $a_p=0$ (from Eq.
\ref{ap}) and the transmission $T_p=T_{13,p}$ (which equals unity
for matching layers, $\epsilon_3=\epsilon_1$).

Note that for non-matching layers ($\epsilon_3\ne \epsilon_1$) there
is a second Brewster angle, $\phi_{B,2}=\arctan(n_3/n_2)$,
corresponding to the wave in layer 2, propagating at the angle
$\phi_2$. One has $a_p=0$ when $\phi_2=\phi_{B,2}$ and the
transmission increases again to $T_{13,p}$. In terms of the incident
wave angle $\phi_1$ this corresponds to $\phi_{B}' =\arcsin (
n_{31}/\sqrt{n_{32}^2+1})$.
In the range of incidence angles $\phi_B
<\phi_1<\phi_{B}'$ (for  $n_{31} >1$)  or $\phi_B'
<\phi_1<\phi_{B}$  (in  case $n_{31} <1$)
one has $a_p<0$ and $T_p>T_{13,p}$.
 Of course, the transmission does not exceed
 unity at any angle.
%%%%%%%%%%%%%%-------------------------
\begin{figure*}[t]
\centerline{\includegraphics[width=0.86\textwidth,clip]{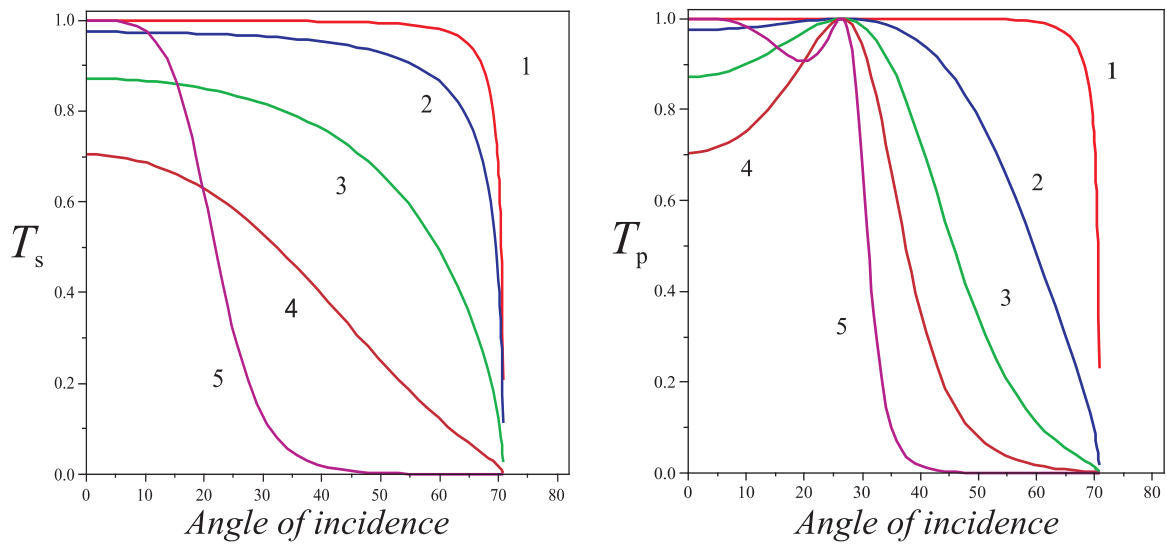}}
\caption[] {(Color online) Angular dependence of the transmission
coefficients $T_s$ and $T_p$ of a thin interlayer with refractive
index $n_2=1.8$ (oxide) for the radiation incident from the first
layer with $n_1=3.6$ (GaAs) and transmitted into the third layer
with  $n_3=3.4$ (InP). Different curves correspond to different
(dimensionless) interlayer thicknesses $\tilde d=d /\lambda$, viz.
(1): $\tilde d=$ 0; (2): $\tilde d=$ 0.02; 3: $\tilde d=$ 0.05; 4:
$\tilde d=$ 0.1, and 5: $\tilde d=1/(2n_2)$.  The left and the right
panels correspond, respectively, to $s$ and $p$ polarizations of
incident light. The TIR angle $\phi_t=$ 30$^\circ$, the Brewster
angles are $\phi_B=26.6^\circ$ and $\phi_B'=26.2^\circ$.}
\label{Fig2}
\end{figure*}
%%%%%%%%%%%%%%%-------------------------
At the FTIR angle, $\phi_1=\phi_t$, both $T_s$  and $T_p$ decrease
at large $d$ as $1/d^2$. Therefore, the widely used approximation,
$T=1$ at $\phi_1< \phi_t$ and $T=0$ $\phi_1 \ge  \phi_t$, works
reasonably well for very thin interlayers (see Fig. \ref{Fig2} for
$\tilde d=0$ and $\tilde d=0.02$) but it does not hold for thicker
films [see Figs. \ref{Fig2} and \ref{Fig3}, for $\tilde d=0.1$ and
$\tilde d=1/(2n_2)$].

Finally, for a thin interlayer  with a finite absorption coefficient
$\alpha$, transmission  $T$ also decrease linearly with $d$.
However, for $ \alpha \lambda^2/(\pi n_2)^2 \ll d$ this is a
negligible effect \cite{Abeles}.

\subsection*{Higher-index overlayer}
As the first example, we consider the transparency of a
semiconductor structure  consisting of two semiconductor layers
(InAs and CdSb) having $n_1=3.5$ and $n_3=4.5$, optically separated
by a chalcogenide glass layer with $n_2=2.6$. This example is close
to parameters of mid-infrared LED  opto-pairs comprising a light
emitting diode and a lens attached to the diode by a layer of
optical glue \cite{LED1}.  Calculated dependencies of the
transmission coefficient on the angle of incidence for two
polarizations and four values of the layer thickness $d$ are shown
in Figs. \ref{FigMLED}, in  the angular range $0 < \phi_1  < \pi
/2$. For this example, the Brewster angles are $\phi_B$=36.6$^\circ$
for the first surface and $\phi_B'$=40$^\circ$ for the second
surface, so that the reflection of $p$-polarized light is strongly
suppressed even in the region close to the angle of TIR from the 1-2
surface, $\phi_t$ = 48$^\circ$.

For thicker interlayers, Figs. \ref{FigMLED} show the growth of
reflection for both polarizations in the region of total internal
reflection. The reflection increases to unity at $\phi_1 =\pi /2$.
For a half-wave plate the transparency is improved only in a range
of angles of incidence smaller than the angle of TIR from the 1-2
interface.

%%%%%%%%%%%%---------------------------
\begin{figure}[b]
\centerline{\includegraphics[width=0.45\textwidth,clip]{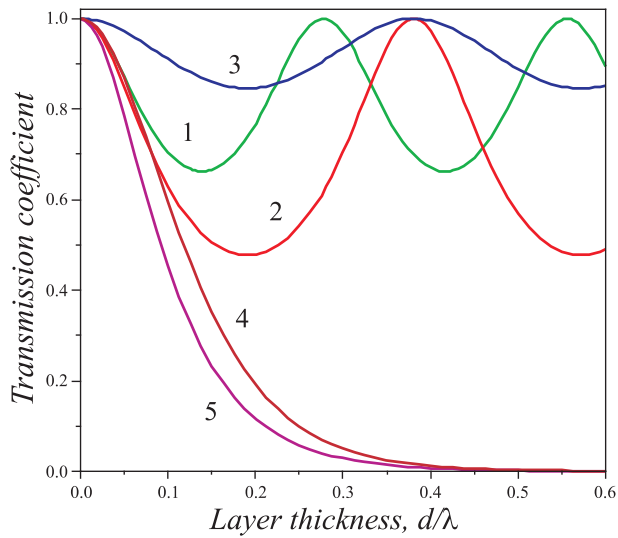}}
\caption[] { (Color online) Dependence of the transmission
coefficients $T_s$ and $T_p$ of a film on the dimensionless layer
thickness $\tilde d=d /\lambda$. The structure parameters are the
same as in Fig. \ref{Fig2} ($\phi_t=30^\circ$). Different curves
correspond to different values of the angle of incidence and
different polarizations of the incident light, viz. (1): $\phi_1=0$,
(2): $\phi_1=20^\circ s$; (3): $\phi_1=20^\circ p$; (4):
$\phi_1=36^\circ s$; (5): $\phi_1=36^\circ p$. } \label{Fig3}
\end{figure}
%%%%%%%%%%%%---------------------------
To ensure that the reflection losses at all angles of incidence
$\phi_1 \le \phi_t$ are less than 10\%, one needs to take $\tilde d
\le 0.05$. For example, in the mid-IR region ($\lambda=$ 3 $\mu$m)
this gives $d \le$ 0.15 \ $\mu$m, which is much more restrictive
than usually expected.
\subsection*{Lower-index overlayer}
Another practical example is a bonded photodetector on top of a
semiconductor scintillator slab with an optical-glue layer in
between. We consider the interlayer transparency in a semiconductor
structure consisting of GaAs layer 1 (emission $\lambda =$ 860 nm)
with the index $n_1=3.6$  and InP photodiode layer 3 of index
$n_3=3.4$. The layers are optically separated by a layer with
$n_2=1.8$ (which seems to be reasonable for the near-IR region).
Figure \ref{Fig2} shows the transmission coefficients $T_{13,s}$ and
$T_{13,p}$, calculated as functions of the incidence angle for
different interlayer thicknesses. Due to the high index contrast,
the TIR angle from the interlayer is only $30^\circ$. A smaller
index of layer 3 ($n_{31}=0.94$) results in additional TIR and
reduces the overall transparency region to $\phi_1 \le 70.8^\circ$.
We note that the reflection losses are quite small for sufficiently
thin films, as well as for $p$-polarized waves at Brewster's angle
and for a ``half-wavelength" thin film. The small reflection is, of
course, owing to low index contrast between layers 1 and 3. Still,
the requirements to the thickness of the interlayer in the
near-infrared region are quite stringent. Even at small angles of
incidence, one should use films with $\tilde d <0.05$ in order to
reduce reflection losses below 10 \%. For $\lambda=$ 860 nm, this
corresponds to $d\le $ 50 nm, which is technologically challenging.
Figure \ref{Fig3} shows the dependence of the transmission
coefficients on the interlayer thickness in this structure for two
polarizations and several angles of incidence, including the normal
incidence and representative angles below and above $\phi_t$. We
clearly see the interference structure that is dependent on the
angle of incidence and almost total transparency at the transmission
maxima for $\phi_1<\phi_t$.
%%%%%%%%-------------------------------
\begin{figure*}[t]
\centerline{\includegraphics[width=0.86\textwidth,clip]{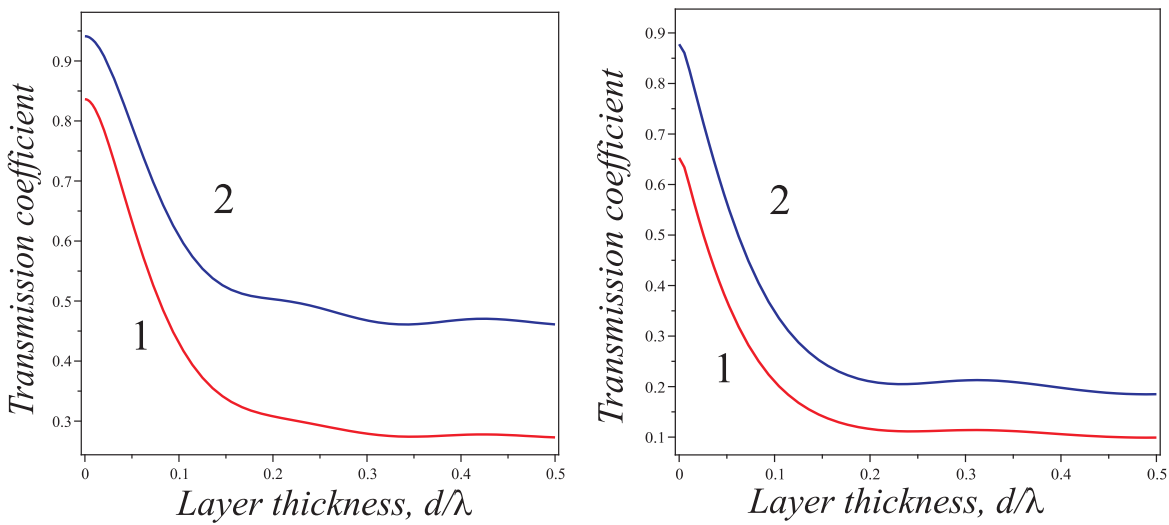}}
\caption[] {(Color online) Average transmission coefficients
$\langle T \rangle$ (denoted as 1) and $\langle T \rangle_{eq}$ (2)
of a film as functions of the film thickness in units of wavelength
$\lambda $ for the structures specified in Figs. \ref{FigMLED} (left
panel) and \ref{Fig2} (right pane).} \label{Fig4}
\end{figure*}
%%%%%%%%%%%%---------------------------
A useful quantity to consider is the {\it average} transmission of
optical power from an isotropic source, discussed in the next
Section. As we shall see below, owing to the angular dependence of
the transmission maxima positions, the average transmission is a
monotonically decreasing function of the interlayer thickness.

\section*{Average transmitted  power }
The average optical power transmitted through the interlayer from a
finite-size non-collimated source depends on the details of the
geometry. Still, one can give meaningful definitions for the average
in two limiting cases relevant to experimental applications (see
Appendix). For radiation from a point-like source (isotropic
non-polarized emission) the appropriate average transmission is of
the form \be \langle T \rangle=\frac{1}{2} \int_0^{\pi/2} [T_s
(\phi_1)+T_p (\phi_1)] \sin \phi_1 d\phi_1 ~. \label{Tavr} \ee
Definition (\ref{Tavr}) corresponds to simple averaging of incident
radiation over the solid angle and can be used to estimate the
external efficiency of LED and scintillators when the emission layer
is transparent, i.e. much thinner than the absorption length.

A different averaging procedure must be used for a distributed
source whose linear dimensions (including the thickness) much exceed
the absorption length. In this case, every small part of the surface
is illuminated by an isotropic flux of radiation coming from the
distance smaller than the absorption length. The common example of
such a source is the quasi-equilibrium (homogeneous) interband
luminescence of an optically excited semiconductor layer when the
excitation region is much larger than the absorption length. The
appropriate average in this case is of the form, \be \langle T
\rangle_{eq}= \int_0^{\pi/2} [T_s (\phi_1)+T_p (\phi_1)] \cos \phi_1
\sin \phi_1 d\phi_1~. \label{Tavre} \ee Definition (\ref{Tavre}) is
implicitly used in the discussion of blackbody radiation (see e.g.
\cite{L&L-StPh}). For non-equilibrium sources, an averaging
procedure similar to (\ref{Tavre}) has been applied, e.g., to the
situation when photon recycling dominates the source properties
\cite{Yablon}. In this case, owing to multiple photon
absorption/re-emission processes, the radiation is widely
distributed over the source layer.

We note that the average (\ref{Tavre}) is normalized (additional
factor of 2 compared to Eq. \ref{Tavr}) so that for a non-reflecting
boundary one has $\langle T \rangle_{eq}=\langle T \rangle =1$. To
compensate for this normalization factor, our expression for the
average power transmission includes an additional factor of 0.5 (see
Appendix). Since the transmission coefficients $T_{13}(\phi_1)$ are
decreasing functions of the incident angle $\phi_1$, one can
generally expect $\langle T \rangle < \langle T \rangle_{eq}$.

An interesting reciprocity relation can be proven for $\langle T
\rangle_{eq}$ (but not for $\langle T \rangle$), viz. \be n_1^2
\langle T \rangle_{eq}|_{1\rightarrow 3}=n_3^2 \langle T
\rangle_{eq}|_{3\rightarrow 1}~. \label{SumRule} \ee This ``sum
rule'' follows from a thermodynamic equilibrium argument (see
Appendix) and holds for any index and arbitrary thickness of the
interlayer. In particular, the averaging in Eq. (\ref {SumRule})
includes the FTIR range of angles, where the interfaces 12 and 23
are totally reflective.

Figure \ref{Fig4} shows the average transmission coefficients
$\langle T \rangle$ and $\langle T \rangle_{eq}$ calculated for
structures with both higher-index (left panel) and lower-index
(right panel) overlayers. We see that for $d/\lambda \ll 1$ the
decrease of the average transmission coefficients with $d$ is almost
linear. The range of quadratic decrease that could be anticipated
from Eq. (\ref{Tfin}) is extremely narrow. The origin of this can be
traced back to analytical expressions. It is evident from Eqs.
(\ref{Tavr},\ref{Tavre}) that the contribution to the average of
small angles of incidence is not decisive: the integrals for
$\langle T \rangle$ and $\langle T \rangle_{eq}$ receive little
contribution from the angles where $T$ is much smaller than unity,
and therefore the average transmission coefficients are very
sensitive to the angular {\it range} where the transmission is high.
With decreasing $d<\lambda$, the range of incidence angles where the
film is highly transparent broadens as $\approx 1/d$, see e.g. Fig.
\ref{Fig2}. As a result, for both polarizations, the decrease of
$\langle T \rangle$ with $d$ becomes linear rather than quadratic.

\section*{Summary and conclusions}
We have analyzed the transparency of low-index intermediate layers
between two higher-index semiconductor layers in the situation when
the angular spread of incident radiation is important. The relevant
applications we have in mind are semiconductor opto-pairs formed
with an optical glue or adhesive.

Transmission of isotropic radiation by thin interlayers with low
refractive index depends on the frustrated total internal reflection
due to the constructive interference of evanescent waves reflected
by the two surfaces of the interlayer. The main contribution to the
average optical power transmission comes from fairly large angles of
incidence. This implies a significant reflection loss even for thin
films of thickness less than the wavelength. For the typical
thicknesses ($\ge 1 \mu$m) of the optical glue interlayers the
losses are high even in the mid-infrared region.

In light of our results, even the best (thinnest) reported oxide
layers used for optical wafer bonding may not be sufficiently thin
for some intended applications. Consider the record-thin 60 nm oxide
bonding layer, recently reported by \cite{Bowers}. For the case of
InP photodiode structure with SiO$_2$ bonding layer of index 1.46
one has the parameter $\tilde d \approx 0.1$ and the reflection is
strong for all angles. To ensure reasonable power transmission
across the bonding layer one must have its thickness $d$ well below
$\lambda/(2\pi n_2) \approx \lambda/12$, which may be quite
challenging technologically.

Special attention has been paid to the angular average of the
optical power transmission, which is shown to possess a universal
form for two practically relevant classes of source layers. We have
found a simple and universal sum rule that must be satisfied by the
angular dependence of the transmission coefficient. The sum rule has
the form of reciprocal relation (\ref{SumRule}) and holds for
three-layer structures with non-absorbing interlayers of arbitrary
thickness and refractive index. It also demonstrates that the
average transmission coefficient of the three-layer structure is
sensitive to the overlayer refractive index relative to that of the
source layer and that structures with larger overlayer index are
preferable.

\section*{Acknowledgements}
We are grateful to Sergey Suchalkin for useful discussions. This
work has been supported by the Domestic Nuclear Detection Office
(DNDO) of the Department of Homeland Security, by the Defense Threat
Reduction Agency (DTRA) through its basic research program, and by
the New York State Office of Science, Technology and Academic
Research (NYSTAR) through the Center for Advanced Sensor Technology
(Sensor CAT) at Stony Brook.

\section*{APPENDIX. Equilibrium radiation and transmitted luminescence}
\setcounter{equation}{0}
\renewcommand \theequation {A.\arabic{equation}}
In general, radiation transmitted through a multi-layered
semiconductor structure may strongly depend on the details of
geometry, especially in case when the linear dimensions of the
emitting region are comparable to the wavelength or the absorption
length. In this situation, one may need detailed modeling that takes
into account the boundary effects. However, there are two limiting
cases when boundary effects are not important. Fortunately, these
limiting cases are precisely those that are most relevant for
applications to semiconductor opto-electronic devices in the visible
and near-infrared range.

The first case corresponds to a transparent emitting layer of large
area and a finite thickness that is substantially smaller than the
absorption length. The second case corresponds to the opposite limit
of interband emission by thick absorbing layers of linear dimensions
much larger than the absorption length. In both limits the
transmission assumes a universal -- albeit different -- form.

Consider first a transparent layer 1 with a point-like isotropic
source of power $P$ located at a distance $z$ from the layer
surface. The normal component of the incident flux through a
circular area $ds=2\pi \rho d\rho$ at the surface (where $\rho=r
\sin \phi_1$) is $dI_{z=0}=P/(4\pi r^2) \cos \phi_1 2\pi \rho d\rho
=P/(4\pi) d\Omega$, where $d\Omega =2 \pi \sin \phi_1 d\phi_1$ is the
solid angle of illumination of the area $ds$. Therefore, in this
case the average transmission coefficient for the power transmitted
to material 3 (through an intermediate layer 2) should be defined as
follows: \be \langle T_{13}\rangle =\int_0^{\pi/2} T_{13}(\phi_1)
\sin \phi_1 d \phi_1~. \label{trans13i} \ee It is important to note
that the transmitted power does not depend on $z$ and hence the same
average transmission coefficient (\ref{trans13i}) determines the
transparency of a unit area of the surface layer for the case when
the source of radiation is distributed in a layer of large area and
finite thickness.

The other limiting case is radiation escape from an absorbing layer.
Consider as an example the total power of equilibrium radiation
emitted at a temperature $T$ by a homogeneous and optically
isotropic material 1. The rate of radiation emission obeys the
detailed balance between the emission and absorption processes
embodied in the van Roosbroeck-Shockley relation \cite{VRS} and is
proportional to the absorption coefficient $\alpha (\omega)$. The
photon density in the unit frequency interval at $\omega$ emitted in
unit volume per unit time is given by \cite{Moss} \be N_\omega
=\frac{n_1^2\omega^2 \alpha(\omega)}{\pi^2 c^2 [\exp(\hbar
\omega/kT)-1]}~. \label{VRSH} \ee On its way out the radiation may
be absorbed and re-emitted many times, but this does not change the
equilibrium photon density. The number of photons reaching  the
surface unit at distance $r$ from the emitting source in volume $dv$
equals \be dI(r)=\frac{1}{4 \pi r^2}\exp (-\alpha r) N_\omega dv~.
\label{Point} \ee At a given incidence angle $\phi_1$, the distance
to the surface $r=z/\cos \phi_1$. Therefore the total photon flux  $
I|_{z=0}$ to the unit surface (at $z=0$) from the region $z>0$  in a
unit solid angle about a {\it fixed} incidence angle $\phi_1$ can be
obtained by integration over $r$, which takes the form \be
I_{\omega}|_{z=0}=\frac {c_1 P_\omega}{4 \pi}\int_0^\infty
\exp\left(-\frac{\alpha z}{\cos \phi_1}\right) \frac{\alpha }{\cos
(\phi_1)}dz \label{BK} \ee where $P_\omega=N_\omega /(\alpha c_1)$
is the equilibrium photon density of thermal radiation and
$c_1=c/n_1$  is the speed of light in the emitting material. Since
the integral in the right-hand side of Eq. (\ref{BK}) equals unity,
the Eq. (\ref{BK}) shows that the equilibrium flux to the surface is
identical to blackbody radiation and depends neither on the
incidence angle nor on the particular shape of $\alpha (\omega)$
(weak emission rate in frequency regions of small absorption is
compensated by the high material transparency at these frequencies).
The energy transmitted to material 3 is given by the integral over
the normal component of the incident flux, multiplied by the
transmission coefficient $T_{13}$, i.e. \be Q_{31}=\int T_{13}\cos
\phi_1 I_{z=0} d \Omega \label{pow}~, \ee where we have suppressed
the frequency index $\omega$, since Eq. (\ref{pow}) is valid for
each frequency individually. For $T_{13}=1$ it gives  a well-known
result $Q_{31}=c_1 P_\omega/4$.  For $T_{13} \ne 1$  we can define
the average transmission coefficient from 1 to 3 by \be \langle
T_{13} \rangle_{eq}= 2 \int_0^{\pi/2} T_{13}(\phi_1)\cos \phi_1 \sin
\phi_1 d\phi_1~. \label{trans13} \ee so that $Q_{31}=\langle T_{13}
\rangle_{eq} c_1 P_\omega/4$.

Note that the shape of Eq. (\ref{pow}) derives from the fact that
the integral in (\ref {BK}) is independent of $\phi_1$, and
therefore it remains valid for a layer of finite thickness, so long
as it much exceeds the absorption length $\alpha^{-1}$. Therefore,
the average (\ref {trans13}) can be used to quantify the surface
transparency in the case of a quasi-equilibrium emission (e.g.
optically excited luminescence) from an absorbing region, so long as
than both its thickness and lateral extent are larger than the
absorption length. For a non-reflective surface $\langle T_{13}
\rangle= \langle T_{13} \rangle_{eq} =1$, while for an interlayer
with finite reflection  $\langle T_{13} \rangle_{eq} \ge  \langle
T_{13} \rangle $, since generally  $T_{13}(\phi_1)$ is a decreasing
function of $\phi_1$.

We remark that the two averaging procedures lead to tangible
differences in power transmission only when the radiation goes
through a large solid angle. The difference is therefore of the
essence for opto-pairs we are interested in, where the refractive
indices $n_1$ and $n_3$ are not vastly disparate. In contrast, for
emission from a semiconductor to low-index media (such as air or
vacuum) the transmission is restricted by the TIR to small angles
and the difference in power transmission is minor \cite{simple}.

Consider now the case of thermal equilibrium between material 1 and
material 3. For the equilibrium to hold, one should have
$Q_{13}\equiv Q_{13}$. This requires that \be n_1^2~\langle
T_{13}\rangle_{eq} =n_3^2~\langle T_{31}\rangle_{eq}
\label{equi13}~. \ee Suppose $n_1 > n_3$. Part of the equilibrium
radiation incident on the interface gets reflected due to the TIR
phenomenon. However, this is exactly compensated by the higher
density of photon states in the higher-index material. This
compensation is ``moderated" by the slower velocity of the energy
flux in the second material, so that the resultant compensating
effect is of the second, and not the third power in $n_{13}$.
Obviously, for a non-equilibrium situation there is no compensation.
But the sum rule expressed by Eq. (\ref{equi13}) remains valid, so
that the calculated transmission coefficient must obey Eq.
(\ref{equi13}).

The sum rule (\ref{equi13}) holds for any planar interface including
any intermediate layer of index $n_2$, so long as there is no
absorption in the intermediate layer. This can be verified by direct
inspection of the integrals  using explicit expressions
({\ref{Tfin}) for both polarizations. To do this, we note that the
transmission coefficients $T_{13}$ and $T_{31}$ can be written as
functions of both the incidence angle and the refraction angle,
subject to Snell's law $n_1 \sin \phi_1 =n_3 \sin \phi_3$, that
holds for both directions of transmission. As an example, we can
write $T_{13,s}$ in the form \be T_{13,s}=\frac{4 n_1 n_3 \cos
\phi_1 \cos \phi_3 }{(n_3 \cos \phi_3 + n_1 \cos \phi_1)^2}
\label{sym13} \ee that makes symmetry between transmission
coefficients 1$\rightarrow$3 and 3$\rightarrow$1 evident. Similarly,
$a_s$ can be written as $a_s=b_{12}~b_{32}$ where \be
b_{ij}=\frac{(n_i^2-n_j^2)}{2n_i \cos \phi_i
 \sqrt{n_j^2-n_i^2 \sin^2 \phi_i}}~,\ee
 and the phase shift as $~\delta = (2\pi d )/\lambda
(n_2^2-n_1^2 \sin^2 \phi_1)^{1/4} (n_2^2-n_3^2 \sin^2
\phi_3)^{1/4}$. Then, one can replace integration  in the left-hand
side of Eq. (\ref{equi13}) over $\phi_1$ by integration over
$\phi_3$ so that $n_1 \cos \phi_1 ~d\phi_1 =n_3 \cos \phi_3
~d\phi_3$ or, with Snell's law, \be n_1^2 \cos \phi_1 \sin \phi_1
~d\phi_1 =n_3^2 \cos \phi_3 \sin \phi_3 ~d\phi_3 ~.\ee Finally, by
changing the integration variable from $\phi_1$ to $\phi_3$ (with an
appropriate change of the integration interval), we arrive at the
direct proof of the sum rule (\ref{equi13}).


\begin{thebibliography}{99}
\bibitem{ref} We remark that Eqs. (\ref{ref12}) evaluated at $\phi_1=0$
produce an opposite sign for the amplitude reflection coefficients
of $s$- and $p$-polarized waves, even though at normal incidence
these waves are identical. It should be noted that reflection
coefficients for $s$ and $p$ polarizations are written down in
(\ref{ref12}) in terms of the electric and the magnetic fields,
respectively. The negative sign for the $p$-polarized wave
originates from the phase difference in the reflected wave.
\bibitem{simple} For emission into vacuum from a semiconductor
with $n \approx 4$, the transmitted fraction of optical power
(restricted only by the total internal reflection) can be estimated
as $1/2n^2$ for both averaging procedures.
\end{thebibliography}
\end{document}